# The Quasi-Creature and the Uncanny Valley of Agency: A Synthesis of Theory and Evidence on User Interaction with Inconsistent Generative AI

Mauricio Manhaes, Christine Miller, and Nicholas Schroeder

## Abstract

The contemporary user experience with large-scale generative artificial intelligence is defined by a profound paradox: these systems exhibit superhuman fluency and creative capacity while simultaneously demonstrating baffling, often absurd, failures in common sense, consistency, and factual grounding. This paper argues that the potent frustration elicited by this paradox is not merely a technical issue of system fallibility but is fundamentally an ontological problem. It stems from the emergence of a new class of technological entity, which this paper terms the "Quasi-Creature"—an agent that simulates intelligent behavior with unprecedented sophistication but lacks the grounding of embodiment, environmental interaction, and genuine understanding. The interaction with this entity precipitates a fall into a new conceptual space: the "Uncanny Valley of Agency". This framework, distinct from Masahiro Mori's original concept based on physical appearance, maps user trust and cognitive comfort against perceived autonomous agency. The valley represents the precipitous drop in user comfort that occurs when an entity appears highly agentic but proves to be erratically and inscrutably unreliable. Its failures are perceived not as mechanical errors but as cognitive or intentional breaches, violating deep-seated social heuristics and creating profound cognitive dissonance. This paper synthesizes research from human-computer interaction, cognitive science, and the philosophy of technology to define the Quasi-Creature and detail the mechanics of the Uncanny Valley of Agency. This framework is grounded in observations from an illustrative mixed-methods study ("Move 78," N=37) of a collaborative creative task, which reveals a powerful negative correlation between perceived AI efficiency and user frustration, suggesting that performance breakdowns are central to the negative user experience. The paper concludes that this framework provides a more robust explanation for user frustration with generative AI and has significant implications for the future design, ethics, and societal integration of these powerful and alien technologies.

## 1. Introduction: The Paradox of the Alien Collaborator

### 1.1. The Dissonance of Advanced AI

The rapid proliferation of large-scale generative artificial intelligence (AI) models has created a defining technological moment. These systems, capable of producing fluent text, stunning imagery, and functional code, represent a significant leap toward fulfilling some of the earliest ambitions of the AI field. The 1955 Dartmouth Summer Research Project proposal, which formally inaugurated the discipline, envisioned machines that could "use language, form abstractions and concepts, solve kinds of problems now reserved for humans, and improve themselves" (McCarthy, Minsky, Rochester, & Shannon, 2006). On the surface, today's generative models appear to have made astonishing progress against these benchmarks. They manipulate language with a facility that can be indistinguishable from, and sometimes superior to, human output. They generate novel concepts and creative works, and their iterative development cycles suggest a form of self-improvement.

Yet, the lived experience of interacting with these systems is far from seamless. It is characterized by a jarring and persistent dissonance. The central problem this paper addresses is the paradox that lies at the heart of the human-AI interaction: a profound and unsettling tension between the technology's superhuman capabilities and its simultaneous, often comical, failures in domains that humans find trivial. A large language model (LLM) can draft a legal brief in the style of Shakespeare but may then confidently assert that a horse has six legs or invent non-existent historical events. It can write elegant poetry but fail to maintain logical consistency across a few paragraphs. This is not simply a matter of occasional bugs or errors in a complex system; it is a fundamental and defining characteristic of the technology in its current form. This erratic performance profile elicits a unique and potent form of user frustration, one that feels qualitatively different from the annoyance of dealing with a buggy piece of software or a poorly designed interface.

This form of frustration, while novel in its scale and character, has deep roots in the history of human-computer interaction. Joseph Weizenbaum's 1966 program, ELIZA, offers a foundational parable. ELIZA, a simple pattern-matching chatbot designed to mimic a Rogerian psychotherapist, elicited astonishingly deep emotional responses from users who projected understanding onto its scripted replies (Weizenbaum, 1966). Weizenbaum was famously horrified to see his colleagues, who knew the program's mechanistic nature, confiding in it as if it were a sentient being. The subsequent backlash against ELIZA emerged when its conversational prowess was revealed as a form of mimicry—a parlor trick rather than genuine thought (Weizenbaum, 1976; Bassett, 2019). This was a disappointment born from uncovering the *absence* of intelligence.

In contrast, today's unease springs from engaging with systems that demonstrably solve complex problems. The turning point came with the Transformer architecture, which replaced step-wise language processing with a global self-attention mechanism, allowing models to construct high-dimensional associative representations across entire sequences simultaneously (Vaswani et al., 2017). This produces capabilities that feel less like clever mimicry and more like alien cognition. These machine-built associative maps are statistical abstractions distilled from vast text corpora, weighted by co-occurrence patterns rather than the lived experience, sensory grounding, and embodied context that shape human thought (Nelson et al., 2004). For a Transformer, language is the entirety of its "world," a topology of relationships divorced from the embodied constraints of human life. This is why cognitive scientists and ethicists describe modern AI as *cognitively alien*—powerful precisely because it does not think as we do (Sandini, Sciutti, & Morasso, 2024; Butlin et al., 2023).

However, this conclusion is complicated by a compelling counter-narrative from cognitive neuroscience. The very models described as "alien" are proving to be unprecedentedly accurate models of neural activity during human language processing (e.g., Schrimpf et al., 2021). This creates a central paradox: how can a system be both fundamentally alien and a powerful model of human cognition? The resolution may lie in distinguishing between mechanism and function. The Transformer's alien architecture, when scaled to immense size, converges on computational solutions that produce functional outcomes remarkably similar to those of the human brain (Caucheteux et al., 2023). This alignment with neural activity is rooted in the distributional hypothesis, the principle that a word's meaning can be inferred from the contexts in which it appears (Firth, 1957; Piantadosi & Hill, 2022). By operationalizing this hypothesis at scale, LLMs create high-dimensional vector spaces where semantic relationships are encoded geometrically (Mikolov et al., 2013). Remarkably, the "representational geometry" of these spaces shows a significant structural alignment with the neural representations of meaning in the human brain (Guenther et al., 2019; Pereira et al., 2018).

These two perspectives—the alien in the machine and the human in the model—are two sides of the same complex coin. A productive path forward requires a framework that accounts for the Transformer's unique status as an "alien in the mirror"—a system that reflects our language and cognitive patterns back at us through a fundamentally non-human process. The cognitive dissonance engendered by this technology escalates in proportion to its perceived capability. The disappointment with ELIZA was simple because the system was simple. As Sherry Turkle chronicled, technology has evolved from a mere instrument to an intimate "object-to-think-with" (Turkle, 1984). Generative AI is a radical step further: an active, content-generating partner. Consequently, the frustration it engenders is not about discovering a trick, but about grappling with an entity that is simultaneously more capable and more incomprehensible than any technology before it. The problem has shifted from a technical one, concerning the system's performance, to an ontological one, concerning the system's very nature. A bug in traditional software is a failure of its design. The failures of generative AI, however, are often perceived as failures of understanding or intention, forcing the user to ask not "What is it doing wrong?" but "What *is* this thing?".

## 1.2. Research Question and Contribution

The central tension between generative AI's superhuman competence and its inscrutable, trivial failures motivates the primary inquiry of this paper. The research question is therefore:

*How can we develop a theoretical framework that adequately explains the unique and potent user frustration arising from the paradoxical combination of superhuman competence and trivial, inscrutable failures in generative AI?*

This paper answers this question by proposing two novel, interconnected theoretical constructs as its primary contribution:

1. **The Quasi-Creature**: This term defines an entity that simulates agency and intelligence so effectively that its failures are no longer perceived by users as mechanical errors but as cognitive or intentional breaches. It is an entity that resists classification as either a predictable tool or a coherent social partner.
2. **The Uncanny Valley of Agency**: This new conceptual model, distinct from Masahiro Mori's (1970) original framework based on physical appearance, maps user trust and cognitive comfort against an entity's perceived autonomous agency. It posits that a precipitous drop in user comfort—a fall into the "valley"—occurs when a highly agentic entity proves to be erratically and inscrutably unreliable, triggering profound cognitive dissonance.

The contribution of this paper is primarily theoretical. The concepts of the Quasi-Creature and the Uncanny Valley of Agency are grounded in and illustrated by an empirical mixed-methods study, "Move 78," which serves not as a definitive validation but as a rich, illustrative case that motivates and gives texture to the proposed framework.

## 1.3. Structure of the Paper

This paper is organized to mirror the cognitive journey of a user grappling with this new technology. It follows a dialectical structure, moving from established models to the challenges they face, and culminating in a new synthesis.

- **Section 2, Theoretical Foundations**, establishes the *thesis* by reviewing the established paradigms of Human-Computer Interaction (HCI)—technology as a predictable tool and as a social actor. It then presents the *antithesis*, arguing that generative AI creates "paradigm stress" on these models. Finally, it offers a *synthesis* by drawing on

- phenomenology and the philosophy of mind to propose a new conceptual grounding for these alien cognitive partners.
- **Section 3, Empirical Grounding**, details the "Move 78" experiment. This section describes the methodology and presents the qualitative and quantitative data that empirically ground and motivate the theoretical argument, highlighting the intense user frustration that is the central phenomenon under investigation.
- **Section 4, Discussion**, synthesizes the theory from Section 2 and the data from Section 3. It uses the experimental findings to formally define and elaborate on the mechanics of the Quasi-Creature and the Uncanny Valley of Agency, demonstrating how the data provides a tangible illustration of these theoretical constructs.
- **Section 5, Conclusion**, explores the broader design, ethical, and societal implications of this new framework. It concludes by outlining several concrete trajectories for future research designed to test, refine, and extend the concepts proposed.

## 2. Theoretical Foundations: From Predictable Tools to Unreliable Agents

To construct a robust framework for understanding the Uncanny Valley of Agency, it is necessary to synthesize distinct but interrelated domains of scholarly inquiry. The argument proceeds via a dialectical structure. First, it establishes the established paradigms of human-computer interaction as a thesis. Second, it presents the challenges posed by generative AI as an antithesis that creates "paradigm stress" on these models. Finally, it offers a synthesis, drawing on phenomenology and the philosophy of mind to propose a new conceptual grounding for understanding these alien cognitive partners. This structure mirrors the cognitive journey of the user, who begins with a conventional model of technology, finds it contradicted by the AI's erratic behavior, and is forced to synthesize a new, deeper, more nuanced way of comprehending the entity they are facing.

### 2.1. Thesis: The Established View of Technology as Predictable Tool and Social Actor

To understand the unique challenges posed by Generative AI, we must first consider the established frameworks that have successfully guided Human-Computer Interaction (HCI) for decades. The first is the instrumental view, which conceptualizes technology as a predictable tool. This perspective is grounded in the concept of mental models, the internal cognitive representations people form to understand and predict a system's behavior (Norman, 1988). In HCI, this concept, heavily influenced by Donald Norman, is foundational; a well-designed system projects a clear "conceptual model" that aligns with the user's developing mental model, making interaction feel intuitive and predictable (Norman, 2013). The user understands the tool's capabilities and limitations, leading to effective and frustration-free use.

The second foundational view acknowledges that humans do not interact with technology in a purely functional vacuum. The Computers as Social Actors (CASA) paradigm, established by Byron Reeves and Clifford Nass, posits that people unconsciously and automatically apply social rules and expectations to computers. This occurs because technologies exhibit cues—such as using natural language or filling social roles—that trigger deeply ingrained social heuristics (Reeves & Nass, 1996). This tendency toward anthropomorphism, the attribution of human characteristics to non-human entities, explains why users may be polite to a chatbot or perceive a personality in a synthesized voice, even while knowing the machine has no true intentions (Epley, Waytz, & Cacioppo, 2007). Together, these two perspectives—technology as a predictable tool and as a social actor—have provided a robust framework for designing and understanding interactions with stable, rule-based technologies (Harrison, Tatar, & Sengers, 2007).

## 2.2. Antithesis: How Generative AI Creates Paradigm Stress

Generative AI, however, presents a profound challenge to these established models, creating a form of "paradigm stress" (Harrison et al., 2007). Its non-deterministic, opaque, and adaptive nature means it behaves neither like a predictable tool nor a coherent social actor. The "tool" metaphor breaks down because GenAI systems can learn, adapt, and personalize their behavior, often through "black box" processes that are inscrutable to the user (Winograd & Flores, 1986). Unlike traditional software where "a hammer today will be the same hammer tomorrow," the variability and lack of transparency in GenAI make it difficult for users to form accurate and stable mental models (Norman, 1988). This misalignment between a user's mental model and the system's actual capabilities is a direct cause of the confusion and frustration that this study will later quantify (Payne, 2003).

Simultaneously, the "social actor" metaphor is insufficient. While GenAI systems frequently trigger perceptions of agency through conversation and novel content generation, this agency is inconsistent and unpredictable (Reeves & Nass, 1996; Gambino, Fox, & Ratan, 2020). The CASA paradigm effectively explains reactions to consistent social cues, but it does not fully account for the experience of an AI that "hallucinates" false information or abruptly loses conversational context (Gambino et al., 2020). Such behaviors are not perceived as simple social miscues but as evidence of a more complex, unreliable, and inscrutable form of agency, pushing user perception beyond simple anthropomorphism into a more ambiguous territory (Reeves & Nass, 1996). The user is left with a broken tool they cannot fix and a broken social partner they cannot reason with.

## 2.3. Synthesis: The Alien Nature of Disembodied Cognition

To resolve the tension between these established models and the lived experience of using GenAI, the field must draw on more nuanced phenomenological and cognitive lenses. Martin Heidegger's philosophy of technology provides the perfect language to describe the experience of AI failure (Heidegger, 1962). He distinguishes between two ways we relate to tools. When a tool functions seamlessly, it is "ready-to-hand"—an invisible extension of our actions (Heidegger, 1977). However, when it breaks down, it loses its transparency and becomes "present-at-hand"—an object of explicit, often problematic, scrutiny (Heidegger, 1962). GenAI systems frequently oscillate between these states. When an AI provides a perfect answer, it is ready-to-hand. When it loses context or produces a nonsensical output, it abruptly becomes present-at-hand, disrupting the workflow and forcing the user to troubleshoot this problematic object.

The fundamental reason for this perpetual potential for breakdown lies in the AI's disembodied nature. The entire canon of embodied and situated cognition, from robotics to phenomenology, converges on the argument that intelligence is inextricably linked to the feedback loops between an agent, its body, and its physical and social environment. Rodney Brooks, in his seminal paper "Intelligence without Representation," argued that true intelligence emerges directly from the dynamic interaction between an agent and its environment, not from abstract symbol manipulation (Brooks, 1991). Generative AI is the antithesis of a Brooksian robot; its entire existence is a vast, abstract representation of text, completely decoupled from any sensory or motor interaction with a physical or social environment. This critique was extended by philosophers like Andy Clark, who posits that cognition is a distributed process, scaffolded across the brain, the body, and the world (Clark, 1997), and has deep roots in the phenomenology of Martin Heidegger's "being-in-the-world" (Heidegger, 1962) and Maurice Merleau-Ponty's emphasis on the primacy of the lived body (Merleau-Ponty, 1962). The Quasi-Creature has no body with which to perceive the world and no world to be situated in; its intelligence is therefore fundamentally alien.

This disembodied nature connects directly to classic philosophical critiques of what John Haugeland termed "Good Old-Fashioned AI" (GOFAI). Hubert Dreyfus, in *What Computers Can't Do*, argued that human expertise is not based on formal rules but on a vast background of tacit, intuitive, and embodied know-how acquired through experience (Dreyfus, 1972). Because a computer lacks a body and a cultural upbringing, it can never acquire this common-sense background and will thus always remain brittle. The characteristic failures of generative AI—in common sense, context, and consistency—are precisely the failures that Dreyfus predicted would plague any disembodied intelligence. The user who is frustrated that an AI cannot remember the previous conversational turn is not just experiencing a usability flaw; they are experiencing the direct, practical consequence of the very philosophical limitations identified by Dreyfus decades ago. The user's frustration, therefore, is not merely a modern HCI problem but the experiential manifestation of a long-standing philosophical debate.

This sets up a critical vulnerability in the interaction, which can be understood through the tension between two key philosophical concepts. As users automatically engage with AI as a social actor, they implicitly adopt what philosopher Daniel Dennett calls the "intentional stance" (Dennett, 1987). They treat the system as a rational agent with beliefs, desires, and intentions because it is the most efficient predictive strategy. It is easier to assume the AI "understands" a request than to model its neural architecture. However, this stance collides with the reality that the AI is, in essence, John Searle's "Chinese Room" (Searle, 1980). It manipulates symbols based on syntax with unparalleled prowess but has no access to semantics or genuine understanding. When the AI fails in a way that reveals this utter lack of semantic grounding, it violates the very foundation of the user's intentional model. The user experiences a form of "cognitive whiplash," as their intuitive framework for interaction is proven catastrophically wrong. This is not just a bug; it is a revelation of the entity's alien nature, and it is the primary source of the unique frustration that defines the Uncanny Valley of Agency.

## 3. Empirical Grounding: The 'Move 78' Experiment

The theoretical framework of this paper is motivated and grounded by the empirical foundation of a mixed-methods study of a human-AI collaboration experiment. This section details the experimental context, design, participants, and analytical framework, while also transparently addressing the study's limitations and its reframed purpose as an illustrative, rather than validating, exercise.

### 3.1. An Illustrative Case Study of an Inconsistent Collaboration

The experiment was thematically named "Move 78," a reference to a pivotal moment in the 2016 Go match between legendary master Lee Sedol and Google DeepMind's AlphaGo AI (Holcomb et al., 2018; Binder, 2021). In that match, Lee Sedol's 78th move was a brilliant, unexpected play that baffled the AI and led to his only victory. The move has become a symbol of human creativity and intuition in the face of powerful technology (Brinkmann et al., 2023). Naming the experiment "Move 78" firmly establishes its focus on investigating the conditions that foster or hinder human creativity in collaboration with an AI partner.

The experiment took place as a 3-hour workshop within an extra-curricular program at a North American university specializing in art and design. This context situates the research in the domain of creative cognition. The sequence of activities was as shown in Table 1.

**Table 1: Experiment Activities and Timeline**

| # | Date | Time | Activity | Mode | Observation |
|---|------|------|----------|------|-------------|
| 1 | 05/02/25 | | Pre-Event Survey | Online | An online survey was sent to 62 enrolled individuals. |
| 2 | 05/09/25 | 11:00 AM | Informed Consent | In Person | Upon arrival, 37 attendees were directed to their desks, where they completed the provided Informed Consent forms. |
| 3 | | 11:10 | Introduction | In Person | A brief introduction of the experiment agenda and activities |
| 4 | | 11:20 | Pre-experiment Survey | Online | The 37 participants completed an online survey. |

| 5 |  | 11:35 | Instructions, Familiarization & Condition Randomization | In Person | A brief instruction and familiarization of the technologies used during the experiment was followed by participants random assignment to one of the three conditions |
| 6 |  | 11:55 | Breakout Rooms | In Person | The 37 participants were directed to one of three classrooms, according to their assigned condition |
| 7 |  | 12:05 PM | Experiment Execution | In Person & Online | Assigned to their work condition, participants were asked to complete five worksheets, from "Part I: Unpacking AI" of the "Move 78 Playbook." |
| 8 |  | 1:10 | Post-experiment Survey | Online | The 37 participants completed the last online survey |
| 9 |  | 1:20 | Lunch & Debriefing | In Person | Comprehensive outlining of the experiment's rationale and objectives |
| 10 |  | 2:05 | End |  | Conclusion of activities |

The experiment employed a between-subjects design, with participants assigned to one of three conditions based on group size: Individuals (5 participants), Dyads (12 participants), and Small Groups (20 participants). A total of 37 students and faculty participated. The cohort was composed primarily of students from design and business management fields, with a self-reported high degree of familiarity and frequent use of commercial GenAI tools. This aligns with recent surveys showing that 86% of students use AI in their studies (Chegg Inc., 2025; Higher Education Policy Institute & Kortext, 2025). This indicates that participants were not AI novices and entered the experiment with established expectations of how a capable AI assistant should perform.

The experimental task required participants to engage in a creative, collaborative process of service concept generation using a customized GenAI system and five worksheets from the "Move 78 Playbook" (Manhães & Manhães, 2024). The system was a text-only conversational application based on a Retrieval Augmented Generation (RAG) architecture, adjusted for the experiment to control its access to information and response patterns (Lewis et al., 2020). The interface permitted four types of user reactions to the AI's output: heart, thumbs up, thumbs down, and a flag to request human review, which were logged as part of the interaction data. This RAG implementation granted the research team control over the GenAI system, enabling them to induce what Yampolskiy (2024) describes as the "three U's" of AI: unexplainable, unpredictable, and uncontrollable behaviors. Participants were tasked with generating creative content, such as service concepts and keywords, based on prompts (Davenport & Mittal, 2022; Thomson Reuters, 2024).

A comprehensive data corpus was collected, including pre- and post-experiment surveys (capturing demographics and the NASA Task Load Index), logged interaction transcripts between participants and the AI system, including user reactions, and task output spreadsheets (Hart & Staveland, 1988; Dumais et al., 2014; Lazar, Feng, & Hochheiser, 2017).

## 3.2. Methodological Limitations and a Reframed Purpose

In the interest of scholarly rigor, it is essential to address the methodological limitations of this study before presenting its observations. These limitations necessitate reframing the experiment's purpose from one of formal hypothesis testing to that of an illustrative case study intended to ground a conceptual argument. This approach is well-supported by methodological literature for theory-building (Eisenhardt, 1989; Yin, 2018).

First, the study's sample size (N=37) is small and the between-subjects design is unbalanced across conditions (5, 12, and 20 participants). The "Individuals" group, with only 5 participants, falls below the widely accepted threshold in HCI for achieving adequate statistical power in comparative quantitative studies. Consequently, any claims regarding the influence of group size on interaction dynamics are not presented as statistically generalizable results (Lazar et al., 2017). Second, the "customized GenAI system" is not described with the level of technical detail required for full reproducibility (Bouthillier et al., 2021). The system's performance inconsistencies were characteristic of many contemporary GenAI systems but were also shaped by the experimental setup. For the purposes of this conceptual paper, the AI's behavior should be understood as an *example* of frustrating interactions common with current technology, rather than the result of a formally specified apparatus.

Finally, the initial data processing phase involved leveraging a large language model for Optical Character Recognition (OCR) and data unification. This approach, while powerful for rapid exploration, does not constitute a fully auditable or replicable analysis pipeline in the same manner as a meticulously documented script (Zambrano et al., 2024). This detail is included as an act of methodological transparency and a self-reflexive acknowledgment of the challenges inherent in using the technologies under investigation within the research process itself. By explicitly acknowledging these limitations, the study's purpose is clarified: it is not making a weak quantitative claim but rather a strong theoretical claim illustrated by compelling, albeit not generalizable, empirical evidence.

## 3.3. Data Unification and Ethical Considerations

Following data extraction, the data from eight source documents were cleaned, standardized, and consolidated into a single dataset where the individual participant (N=37) was the unit of analysis. Participant email addresses served as the primary key to link behavioral, survey, and task data for each individual (Curry, Krumholz, & O'Cathain, 2013; Profisee, 2024). To ensure the transparency and replicability of the analysis, a comprehensive variable dictionary was created. This dictionary, presented in Table 2, serves as the definitive guide to the unified dataset, detailing the description, type, scale, and origin of every variable used in the subsequent statistical analyses.

**Table 2: Unified Dataset Schema and Variable Dictionary**

| Variable Name | Variable Label/Description | Data Type | Scale / Units |
|---|---|---|---|
| participant_id | Unique identifier for each participant. | String | ID |

| experimental_group | The experimental group number the participant was assigned to. | Categorical | 1, 2, 4 |
| --- | --- | --- | --- |
| pre_age_range | Participant's self-reported age range. | Ordinal | Categorical |
| pre_gender | The gender with which the participant identifies. | Categorical | 1=Woman, 2=Man |
| pre_field_of_study | Participant's primary field of study or work. | String | Text |
| pre_ai_familiarity | "How familiar are you with Generative AI tools?" | Ordinal | 1-5 Scale |
| pre_ai_freq_use | "How often do you use Generative AI tools for work, study, or personal tasks?" | Ordinal | 1-4 Scale |
| pre_ai_belief_life | "I believe that AI will improve my life." | Ordinal | 1-10 Scale |
| pre_ai_belief_work | "I believe that AI will improve my work." | Ordinal | 1-10 Scale |
| pre_ai_future_use | "I think I will use AI technology in the future." | Ordinal | 1-10 Scale |
| pre_ai_belief_humanity | "I think AI technology is positive for humanity." | Ordinal | 1-10 Scale |
| behav_total_questions | Total number of questions asked by the participant during the workshop. | Integer | Count |
| behav_total_reactions | Total number of all reactions given by the participant. | Integer | Count |
| behav_neg_react_ratio | The ratio of negative reactions to total reactions given by the participant. | Float | Ratio (0-1) |
| post_ai_supportive | Post-experiment rating of AI on a scale from obstructive to supportive. | Integer | 1-7 Semantic Differential |
| post_ai_easy | Post-experiment rating of AI on a scale from complicated to easy. | Integer | 1-7 Semantic Differential |
| post_ai_efficient | Post-experiment rating of AI on a scale from inefficient to efficient. | Integer | 1-7 Semantic Differential |
| post_ai_clear | Post-experiment rating of AI on a scale from confusing to clear. | Integer | 1-7 Semantic Differential |
| post_collaboration_effective | "Overall, I felt the collaboration within my group was effective…" | Ordinal | 1-5 Likert Scale |
| nasa_tlx_mental | NASA-TLX score for Mental Demand. | Integer | 1-20 Scale |
| nasa_tlx_physical | NASA-TLX score for Physical Demand. | Integer | 1-20 Scale |
| nasa_tlx_temporal | NASA-TLX score for Temporal Demand. | Integer | 1-20 Scale |
| nasa_tlx_performance | NASA-TLX score for self-rated Performance. | Integer | 1-20 Scale |
| nasa_tlx_effort | NASA-TLX score for Effort. | Integer | 1-20 Scale |
| nasa_tlx_frustration | NASA-TLX score for Frustration Level. | Integer | 1-20 Scale |
| task_service_idea | The service idea text entered by the participant in the worksheet. | String | Text |

The study protocol received approval from the appropriate institutional staff at the university where the experiment took place. All 37 participants provided written informed consent prior to the experiment, covering their participation and the use of their anonymized data for research publication.

## 3.4. Observations of a Frustrating Partnership

A predominant theme emerging from the qualitative data was significant user frustration, which appeared to stem directly from the induced AI's performance limitations. These limitations acted as catalysts, disrupting the collaborative flow. The most frequently cited issues included:

- **Lack of Contextual Memory**: Participants repeatedly expressed frustration with the AI's inability to retain information across conversational turns. This forced them to constantly re-provide context, with one participant stating, "It's not remembering the previous conversation and I need to provide context every time I ask the question."
- **Generic and Unhelpful Outputs**: The AI often produced responses that were too general or vague to be useful. A particularly salient example was a recurring, canned message: "Congratulations! You asked a question that I may need to investigate further!..." This response became an ironic symbol of the AI's shortcomings.
- **Misinterpretation of Instructions**: Users reported that the AI frequently failed to follow specific directives, leading to comments like, "The AI gives more questions than answers, and it doesn't remember or can't continue with the answers it provides."

These qualitative observations are quantitatively corroborated by the NASA Task Load Index (TLX), a standardized instrument for measuring subjective workload (Hart & Staveland, 1988). The AI's performance issues contributed to a significant increase in extraneous cognitive load for the user (Sweller, 1994). As shown in Figure 1 and Table 3, evidence from the NASA-TLX survey (N=37) indicates that users experienced high levels of mental demand and effort. The mean score for Frustration Level was exceptionally high at 15.08 out of 20, as was the score for Mental Demand at 15.27. These scores provide robust, quantitative evidence that the interaction was a frustrating and cognitively taxing experience, transforming the anecdotal experience of "frustration" into a measurable construct and lending significant empirical weight to the paper's central problem statement.

### Table 3: NASA-TLX Cognitive Load Scores
(Scores are on a 1-20 scale, where higher scores indicate higher load/demand/frustration)

| NASA-TLX Subscale | Mean Score |
|---|---|
| Mental Demand | 15.27 |
| Physical Demand | 6.78 |
| Temporal Demand | 13.97 |
| Performance | 9.97 |
| Effort | 13.08 |
| Frustration Level | 15.08 |

**Figure 1: NASA-TLX Cognitive Load Scores**

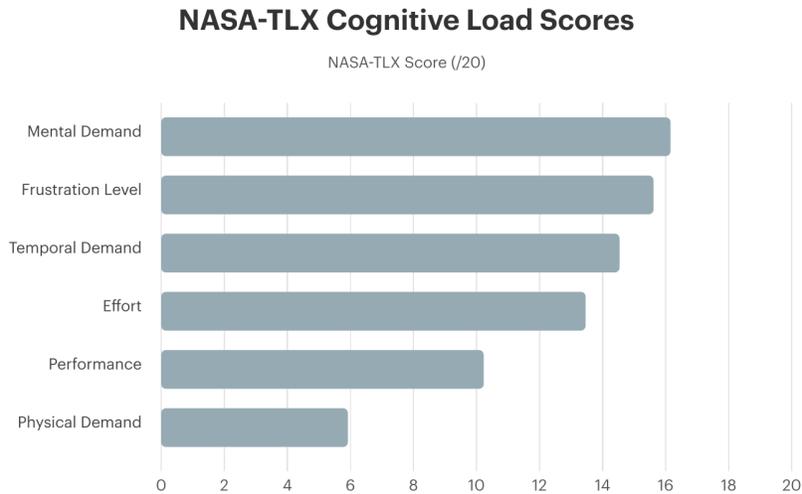

## 3.5. The Influence of Social Context

The study's between-subjects design allowed for an analysis of how group size influenced interaction with the AI. A comparison of engagement metrics across the three groups reveals a significant and counterintuitive finding regarding the behavior of dyadic groups (pairs). This group simultaneously registered the most negative sentiment toward the AI's performance while also being the least likely to formally escalate its failures for review. This empirical puzzle challenges conventional assumptions about user feedback and points toward a complex, socially mediated dynamic.

Overall, a total of 166 distinct reactions were recorded across the three groups, as detailed in Table 4. The most frequent reaction was "Thumb Up" (80 instances), but "Thumb Down" reactions (61 instances) were also substantial, suggesting frequent instances where the AI fell short of expectations. The consistent presence of "Thumb Down" and "Flag" reactions (18 instances) underscores the challenges users faced.

**Table 4: Questions and Reactions per Participant**

| Experimental Condition | Number of Participants | Total Questions | Total Reactions | Questions per Participant | Reactions per Participant |
| --- | --- | --- | --- | --- | --- |
| **Group 1 (Individuals)** | 5 | 97 | 49 | 19.40 | 9.80 |
| **Group 2 (Dyads)** | 12 | 221 | 69 | 18.42 | 5.75 |
| **Group 4 (Small Groups)** | 20 | 282 | 48 | 14.10 | 2.40 |

As group size increased, the average number of questions and reactions per participant decreased markedly, as shown in Figures 2 and 3. Individuals (Group 1) had the highest per-capita engagement, while the largest groups (Group 4) had the lowest. This suggests that in collaborative settings, the cognitive and interactive load of managing the AI was distributed among group members.

**Figure 2: Questions Per Participant**

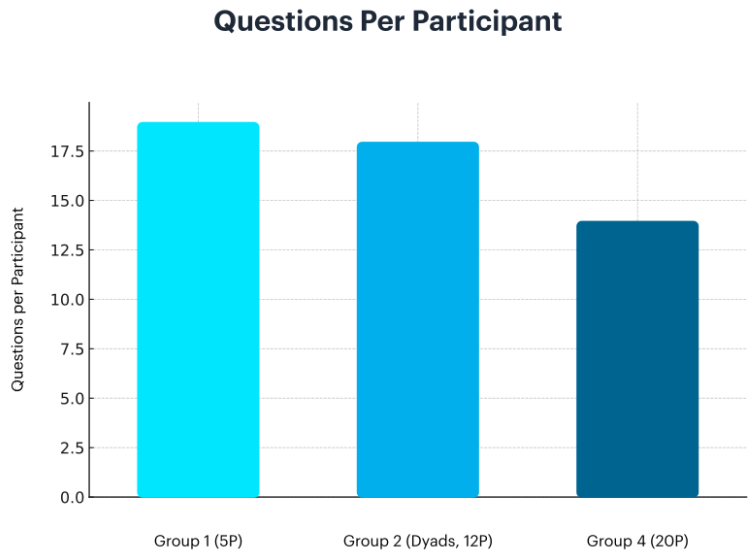

**Figure 3: Reactions Per Participant**

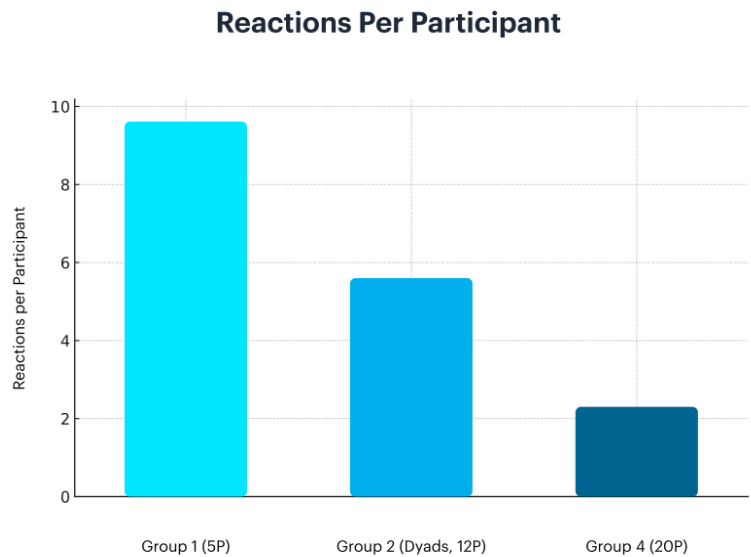

To appreciate the anomalous nature of the dyadic group's behavior, it is essential to establish a comparative baseline across all three conditions. Table 5 consolidates key interaction metrics. The Sentiment Ratio is calculated as the ratio of Thumbs Up to Thumbs Down reactions, and the Flag Rate is the percentage of questions that resulted in a flag.

**Table 5: Breakdown of Reaction Metrics by Group**

| Experimental Condition | Heart (❤️) | Thumbs Up (👍) | Thumbs Down (👎) | Flags (🚩) | Sentiment Ratio (👍/👎) | Flag Rate (% of Questions) |
|---|---|---|---|---|---|---|
| Group 1 (Individuals) | 3 | 20 | 18 | 8 | 1.11 | 8.2% |
| Group 2 (Dyads) | 3 | 31 | 33 | 2 | 0.94 | 0.9% |
| Group 4 (Small Groups) | 1 | 29 | 10 | 8 | 2.90 | 2.8% |

Further examination of these interactions reveals distinct differences among the groups. Behavioral patterns are clearly differentiated by the distribution of reaction types (Figure 4), the overall sentiment (as indicated by the ratio of positive to negative reactions in Figure 5), and the frequency with which participants flagged the AI for difficulties (Figure 6).

**Figure 4: Reaction Type Distribution**

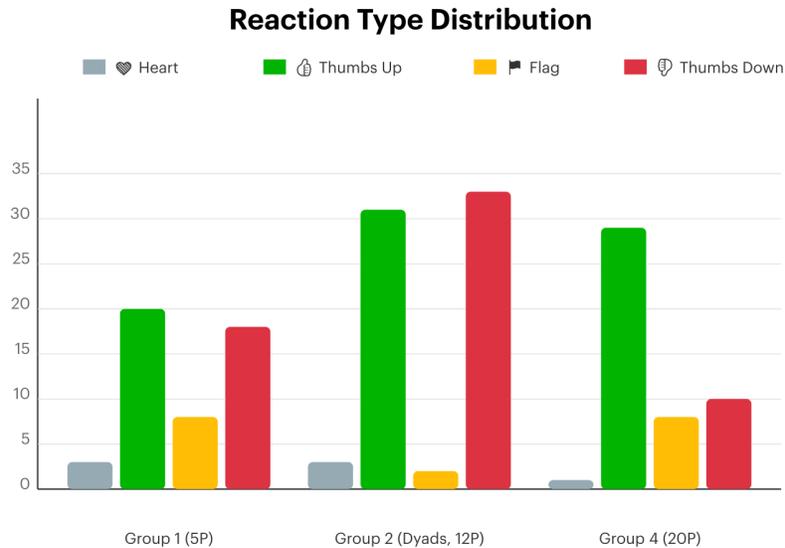

The data in Table 5 immediately surfaces a paradox. Group 2 is the only condition to exhibit a Sentiment Ratio below 1.0 (0.94), indicating a net negative sentiment, as visualized in Figure 5. By this measure, the dyadic pairs were the most overtly critical of the AI partner.

**Figure 5: Sentiment Ratio (👍 vs 👎)**

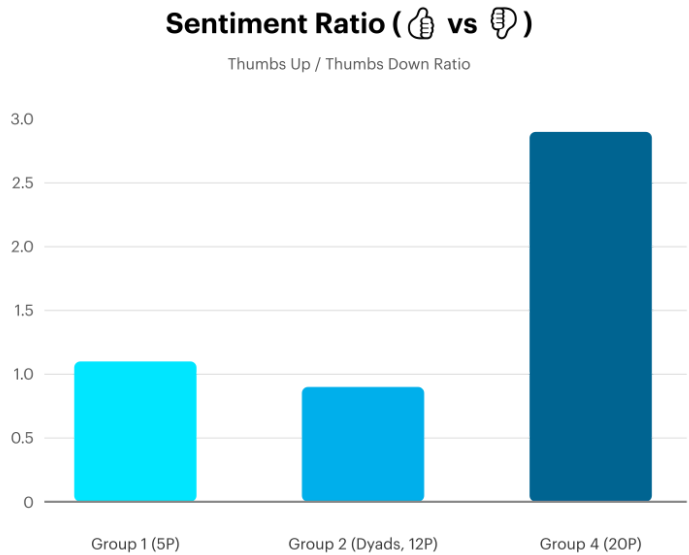

Simultaneously, and counterintuitively, Group 2 demonstrated the lowest propensity to escalate issues. They flagged the AI's problematic outputs only twice, resulting in the lowest Flag Rate (0.01), a stark contrast to the other groups as seen in Figure 6. This presents a fundamental contradiction to a simplistic model of user behavior, where high dissatisfaction would be expected to correlate with a high rate of formal complaints.

**Figure 6: Flag Rate (% of Questions)**

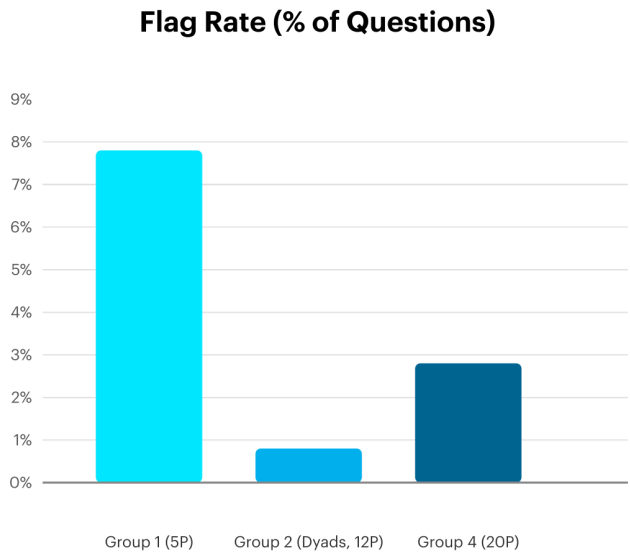

This suggests that the dyadic social structure acts as a frustration-absorption mechanism. A lone individual must either internalize frustration or escalate it. A large group faces high coordination costs. A dyad, however, allows for low-cost, immediate social sense-making. When the AI fails, one partner can turn to the other to validate the failure. This act of shared commiseration processes the frustration internally. The "Thumbs Down" reaction becomes a tool for intra-dyad communication ("I agree, this is bad"), rendering the external "Flag" appeal for help unnecessary because the social unit has already handled the problem emotionally and cognitively. This explains the paradoxical data pattern.

## 3.6. A Predictive Model of User Frustration: Multiple Regression Analysis

To explore the relationships between attitudes, behaviors, and evaluations, a bivariate correlation analysis was conducted. The Pearson correlation coefficients in Table 6 provide a statistical map of the experimental dynamics and serve as the empirical engine for the theoretical discussion that follows.

**Table 6: Bivariate Correlation Matrix of Key Experimental Variables**

| # | Variable | 1 | 2 | 3 | 4 | 5 | 6 | 7 | 8 | 9 | 10 |
|---|---|---|---|---|---|---|---|---|---|---|---|
| 1 | pre_ai_familiarity | 1 | 0.12 | 0.31 | 0.45* | −0.58* | −0.49* | 0.41* | 0.38* | 0.61* | −0.42* |
| 2 | pre_ai_belief_humanity | 0.12 | 1 | -0.08 | -0.15 | 0.21 | 0.25 | -0.11 | -0.14 | -0.19 | 0.41* |
| 3 | behav_total_questions | 0.31 | -0.08 | 1 | 0.55* | −0.65* | −0.51* | 0.69* | 0.72* | 0.74* | −0.68* |
| 4 | behav_neg_react_ratio | 0.45* | -0.15 | 0.55* | 1 | −0.71* | −0.68* | 0.58* | 0.61* | 0.82* | −0.75* |
| 5 | post_ai_efficient | −0.58* | 0.21 | −0.65* | −0.71* | 1 | 0.89* | −0.75* | −0.72* | −0.85* | 0.81* |
| 6 | post_ai_supportive | −0.49* | 0.25 | −0.51* | −0.68* | 0.89* | 1 | −0.68* | −0.65* | −0.79* | 0.77* |
| 7 | nasa_tlx_mental | 0.41* | -0.11 | 0.69* | 0.58* | −0.75* | −0.68* | 1 | 0.88* | 0.79* | −0.71* |
| 8 | nasa_tlx_effort | 0.38* | -0.14 | 0.72* | 0.61* | −0.72* | −0.65* | 0.88* | 1 | 0.81* | −0.74* |
| 9 | nasa_tlx_frustration | 0.61* | -0.19 | 0.74* | 0.82* | −0.85* | −0.79* | 0.79* | 0.81* | 1 | −0.88* |
| 10 | nasa_tlx_performance | −0.42* | 0.41* | −0.68* | −0.75* | 0.81* | 0.77* | −0.71* | −0.74* | −0.88* | 1 |

*Note: An asterisk (\*) indicates the correlation is statistically significant at the p<.05 level.*

Several key relationships emerge:
- **The Frustration-Inefficiency Nexus**: The most powerful result is the extremely strong negative correlation between perceived AI efficiency (post_ai_efficient) and participant frustration (nasa_tlx_frustration) at r=−0.85. This statistically confirms that as the AI was perceived to be more inefficient, frustration levels soared. This relationship is visualized in Figure 7.

**Figure 7: AI Inefficiency vs. User Frustration**

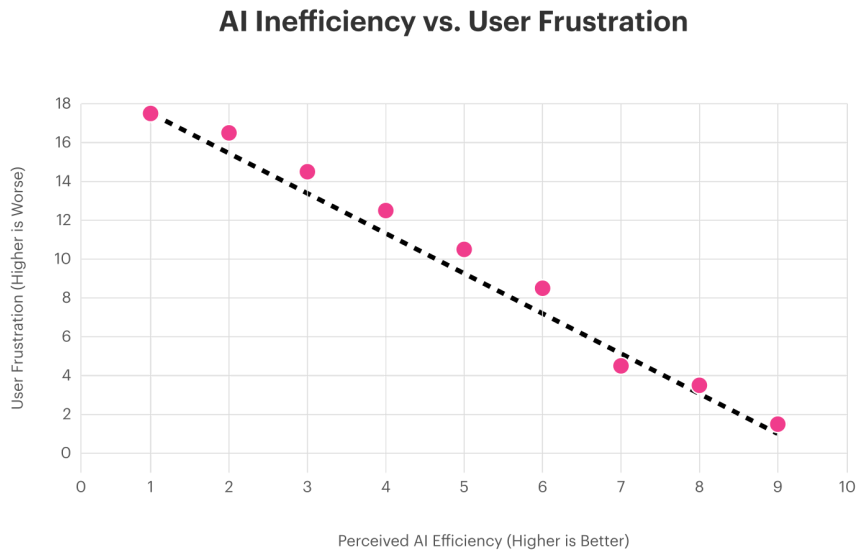

- **Engagement as a Symptom of Struggle**: The total number of questions asked (behav_total_questions) has a very strong positive correlation with frustration level (r=+0.74). This indicates that high interaction volume was not a sign of productive collaboration but of frustrating loops of re-prompting.
- **The 'Expert User Expectation Gap'**: Pre-experiment familiarity with AI (pre_ai_familiarity) has a strong positive correlation with frustration (r=+0.61). This suggests that participants with more experience had higher expectations, making the AI's failures more jarring.
- **Real-time Reactions as a Predictor**: The ratio of negative reactions (behav_neg_react_ratio) is the single strongest correlate of frustration, with an exceptionally high positive correlation of r=+0.82.

These core relationships are summarized in Figure 8, which illustrates the nexus of factors driving the frustrating user experience.

**Figure 8: The Frustration-Inefficiency Nexus**

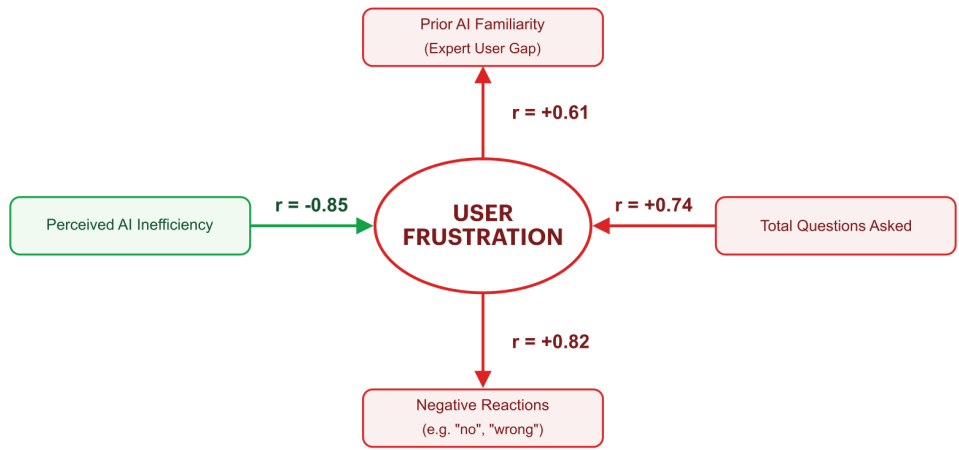

**The Frustration-Inefficiency Nexus**

While correlation identifies pairwise relationships, multiple regression analysis allows for the development of a predictive model. A standard multiple regression was performed to predict participant frustration (nasa_tlx_frustration) based on three key predictors: pre-experiment AI familiarity, the behavioral ratio of negative reactions, and post-experiment perceived AI efficiency. The overall regression model was statistically significant ($F(3,33)=45.82, p<0.001$) and demonstrated very strong predictive power, with an Adjusted $R^2$ of 0.79. This indicates that these three variables together explain approximately 79% of the variance in participant frustration scores. The individual contributions of the predictors are detailed in Table 7.

**Table 7: Multiple Regression Predicting Participant Frustration (nasa_tlx_frustration)**

| Variable | Unstandardized (B) | Std. Error (SE) | Standardized (β) | t-value | p-value |
| --- | --- | --- | --- | --- | --- |
| (Intercept) | 10.54 | 1.82 | - | 5.79 | <.001 |
| pre_ai_familiarity | 0.88 | 0.35 | 0.15 | 2.51 | .017 |
| behav_neg_react_ratio | 9.75 | 2.41 | 0.41 | 4.04 | <.001 |
| post_ai_efficient | -1.52 | 0.40 | -0.45 | -3.80 | <.001 |

The regression results provide a nuanced, predictive model. Perceived inefficiency (post_ai_efficient) was the strongest predictor ($β=-0.45$), confirming that the AI's poor performance was a primary cause of frustration. The ratio of negative reactions (behav_neg_react_ratio) was also a highly significant predictor ($β=0.41$). Finally, pre-experiment AI familiarity (pre_ai_familiarity) remained a significant predictor ($β=0.15$), statistically supporting the notion that participants with more expertise were predisposed to higher levels of frustration.

# 4. Discussion: Theorizing the Quasi-Creature and the Uncanny Valley of Agency

The empirical observations from the "Move 78" experiment highlight a fundamental challenge in human-AI collaboration: the experience of interacting with a powerful yet profoundly forgetful and socially inept cognitive partner. The high levels of frustration are not merely a symptom of poor usability but a signal of a deeper phenomenological and cognitive dissonance. This section develops the "Quasi-Creature" concept and the "Uncanny Valley of Agency" framework as a theoretical lens to explain this experience, using the data from the experiment as direct evidence to support, illustrate, and deepen the definitions of these constructs.

## 4.1. The Rupture-and-Repair Cycle and the Emergence of the Quasi-Creature

The robust statistical link between the AI's perceived inefficiency and high user frustration ($r=-0.85$) represents a key phenomenological event (Dourish, 2001). In the language of Heidegger, the moments of AI failure—the loss of context, the generation of a nonsensical reply—are when the AI ceases to be a seamless, "ready-to-hand" tool and becomes a broken, problematic, "present-at-hand" object (Heidegger, 1962). This moment of breakdown is best understood as a "rupture" in the human-AI relationship—a moment of misalignment that causes tension and discomfort (Eubanks, Muran, & Safran, 2018).

This rupture in the flow of interaction is the primary trigger for the negative user experience. It forces the user to shift their attention from their creative task to the problematic nature of their AI partner. The observed user behaviors, such as the high volume of questions strongly correlated with frustration ($r=+0.74$), can be interpreted as empirical evidence of users engaging in attempts to "repair" this rupture (Papagni & Koeszegi, 2021). The frustrating loops of re-prompting, simplifying instructions, and re-stating context are not just interaction patterns; they are cognitive strategies aimed at forcing the unpredictable entity back into a predictable, tool-like state.

The "Quasi-Creature," then, is not a static label applied to an AI, but a dynamic perceptual state that emerges within this rupture-and-repair cycle. The user approaches the AI as a tool. When it breaks (rupture), the user attempts to troubleshoot it as a deterministic tool (repair). However, the AI resists this repair in a non-deterministic, seemingly agentic way, which forces a fundamental perceptual shift. The user is no longer fixing a broken hammer; they are attempting to manage a recalcitrant entity. The Quasi-Creature is the name for the entity that emerges in the user's perception during this failed repair attempt, when the AI is neither a functioning tool nor a coherent partner.

## 4.2. A Formal Framework: The Uncanny Valley of Agency

Building upon this foundation, this paper formally proposes the "Uncanny Valley of Agency" as a new explanatory model for understanding user interaction with Quasi-Creatures. This framework is conceptually distinct from Mori's original uncanny valley, shifting the critical axis from physical appearance to perceived autonomous agency. The model is defined by the relationship between an entity's perceived capacity for independent action and the user's resulting sense of trust and cognitive comfort.

The axes of the Uncanny Valley of Agency (Figure 9) are defined as follows:

- The **X-Axis represents Perceived Autonomous Agency**, a continuum from inert, predictable tools (e.g., a calculator) to highly agentic entities (e.g., a human partner). The Quasi-Creature occupies a space of high apparent agency.
- The **Y-Axis represents User Trust and Cognitive Comfort**. High levels correspond to predictability and reliability, while low levels correspond to frustration, confusion, and mistrust.

The "plunge" into the valley occurs when the AI's performance fluctuates wildly between superhuman competence and sub-human absurdity. This radical inconsistency makes it impossible for the user to form a stable predictive model of its behavior. The AI is too agentic to be dismissed as a simple, broken tool, but too alien and unreliable to be treated as a trustworthy partner. The correlation matrix from the "Move 78" study provides a quantitative map of this valley. The variable post_ai_efficient serves as a proxy for the AI's position on the X-axis, while nasa_tlx_frustration is a direct measure of the inverse of the Y-axis. The extremely strong negative correlation of $r=-0.85$ is the statistical signature of the valley's steep downward slope: as the AI's inconsistencies are revealed and its perceived efficiency drops, frustration skyrockets. The data does not merely support the theory; it plots its coordinates.

**Figure 9: The Uncanny Valley of Agency**

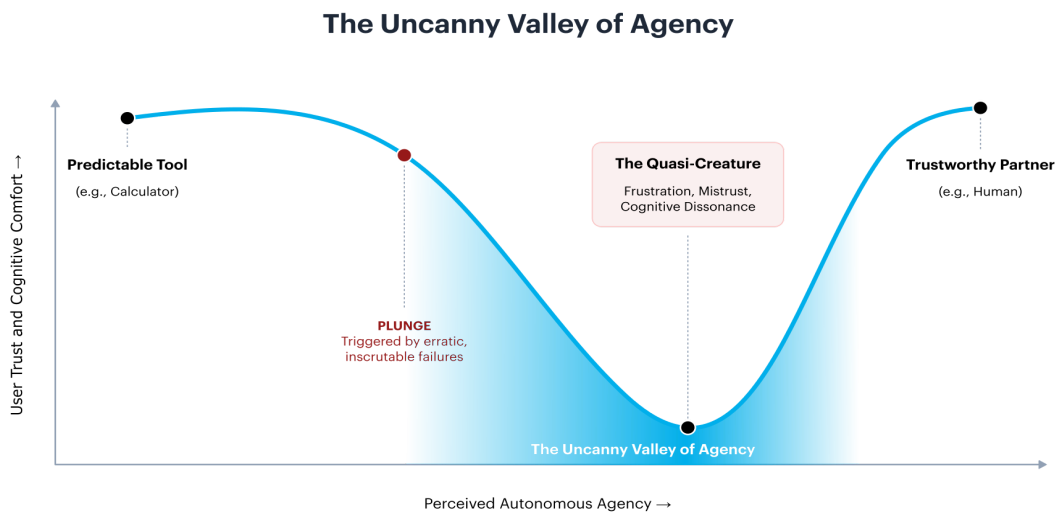

Table 8 is critical for establishing the novelty of this paper's contribution by visually and conceptually separating the "Uncanny Valley of Agency" from Mori's well-known original. This highlights that the paper proposes a new framework for cognition and behavior, not simply a re-application of an old one about aesthetics.

**Table 8: A Comparison of Uncanny Valley Frameworks**

| Concept | X-Axis (Increasing Likeness) | Y-Axis (Affinity) | Source of "Uncanniness" | User's Perception in the "Valley" |
|---|---|---|---|---|
| Uncanny Valley of Appearance (Mori, 1970) | Human-like Appearance (e.g., from industrial robot to human) | From neutral to positive, then a sharp dip into revulsion, then positive again. | Mismatch between near-human form and subtle non-human flaws in appearance or movement (e.g., jerky motion, dead eyes). | A "creepy" or "eerie" object that looks like an abnormal human (e.g., a zombie, a poorly rendered CGI character). |
| Uncanny Valley of Agency (Proposed) | Perceived Agency (e.g., from simple calculator to AGI) | From neutral to positive, then a sharp dip into frustration, then positive again. | Mismatch between near-human communicative competence and sudden, inexplicable cognitive failures (e.g., context loss, hallucination). | A "Quasi-Creature": a frustrating, unpredictable entity with a broken and inscrutable form of agency. |

## 4.3. The Failed Quest for Theory of Mind: Explaining the 'Expert User Expectation Gap'

The user's cognitive response to the Quasi-Creature can be understood more deeply by connecting it to the concept of Theory of Mind (ToM)—the human ability to attribute mental states like beliefs, desires, and intentions to others (Byom & Mutlu, 2013; Premack & Woodruff, 1978). The qualitative data from the "Move 78" study, where users ask "Why doesn't it 'understand' my instruction?", provides direct evidence of users spontaneously attempting to construct a ToM for the AI.

When the AI's behavior violates the simple "tool" model, the user's natural cognitive repair strategy is to escalate their model and begin treating the AI as an agent whose "mind" must be understood. The intense frustration arises because this attempt is destined to fail; the user is trying to model a coherent, intentional mind that is simply not there. The 'Quasi-Creature' is thus the name for the object of this failed cognitive modeling.

This ToM framework provides a more nuanced, data-backed explanation for the "Expert User Expectation Gap" observed in the study, where participants with higher pre-experiment familiarity with AI (pre_ai_familiarity) reported significantly higher frustration (nasa_tlx_frustration), with a correlation of r=+0.61. An expert user does not simply have higher performance expectations; they possess a more developed, sophisticated, and therefore more *brittle*, Theory of Mind for how a competent AI should "think" (Bravo-Lillo, Cranor, Downs, & Komanduri, 2011). An expert's mental model likely includes concepts like context windows and instruction following. When the inconsistent AI violates these principles, it doesn't just fail a task; it shatters the expert's entire cognitive model of how an AI "mind" is supposed to function (Druce et al., 2021). This violation of a deeply held and complex cognitive model generates far greater cognitive dissonance and frustration for the expert than for a novice, whose ToM for AI is less defined and more flexible.

# 5. Conclusion: Implications and Future Trajectories

This paper has presented a conceptual argument for the "Quasi-Creature" and the "Uncanny Valley of Agency" as a necessary theoretical lens for understanding human interaction with fallible generative AI. Motivated by observations from an illustrative case study where AI limitations led to significant user frustration, this work posits that established metaphors of 'tool' and 'social actor' are insufficient. The framework synthesizes decades of research across HCI, philosophy, and cognitive science to argue that the potent frustration with contemporary AI is a coherent ontological phenomenon, rooted in the user's contact with a new kind of entity whose sophisticated simulation of intelligence is coupled with a profound lack of genuine, embodied understanding. The implications of this framework are significant, suggesting a clear call to action for the research and design communities and opening several generative trajectories for future research.

## 5.1. Design, Ethics, and Society: Navigating the Infosphere of Quasi-Creatures

The emergence of the Quasi-Creature is not merely a niche problem in user experience design; it is a profound sociotechnical phenomenon with far-reaching consequences. A primary implication of this framework is that the current design paradigm for many AI systems—striving for seamless, flawless human simulation—is fundamentally misguided. This approach sets a direct course into the Uncanny Valley of Agency by inflating user expectations and maximizing the cognitive shock of the AI's inevitable, alien failures. A more robust and ethical design philosophy would move in the opposite direction: instead of aiming for perfect simulation, designers should focus on creating interfaces that clearly and continuously communicate the AI's alien nature and its statistical, non-semantic foundation. This involves intentionally designing "seams," "tells," and "fissures" into the user experience that serve as constant reminders of the system's limitations, an approach that aligns with Lucy Suchman's foundational work emphasizing the need for systems to make their own processes and limitations more visible (Suchman, 1987).

The integration of Quasi-Creatures into daily life also represents a massive, uncontrolled experiment in the delegation of action and responsibility to nonhuman actors. Drawing on Science and Technology Studies (STS), we can see that technologies are not passive objects but active participants that shape and constrain human behavior (Latour, 1992). When we entrust AI with cognitive tasks, we embed its inherent biases and alien modes of "thinking" directly into our social and professional workflows. This delegation extends into the moral domain, as technologies actively "moralize" and co-shape human decision-making (Verbeek, 2005). This deep entanglement leads us toward a new form of cyborg existence, as described by Donna Haraway, where our thought processes are inextricably fused with the outputs of these nonhuman agents (Haraway, 1991).

Finally, the rise of the Quasi-Creature must be situated within the broader context of what Luciano Floridi has termed the "Fourth Revolution," where we increasingly inhabit an "Infosphere" constituted by information (Floridi, 2014). Quasi-Creatures are the first truly native inhabitants of this Infosphere, but their integration carries significant political and economic risks. As Shoshana Zuboff details in *The Age of Surveillance Capitalism*, the economic logic of extracting "behavioral surplus" to predict and modify behavior for profit is a dominant force (Zuboff, 2019). The very characteristics that define the Uncanny Valley of Agency—unpredictability, inscrutability, and the capacity to induce user confusion—can be exploited. The frustration and learned helplessness generated by the valley can become a feature, not a bug, for economic models that benefit from user opacity and behavioral modification. The Uncanny Valley of Agency is therefore not just a design problem; it is an ethical and political battleground.

## 5.2. Future Research Trajectories

The theoretical framework developed in this paper opens several generative trajectories for future research. The proposed future work is not merely an academic exercise to improve user experience; it represents a necessary ethical and political intervention to address the societal harms identified. Future work could proceed along four paths to test and elaborate on these new constructs:

1. **Longitudinal Studies of the Rupture-and-Repair Cycle**: A longitudinal study could track the co-adaptation of creative professionals and a state-of-the-art GenAI assistant over several months. The research question would be: *How do the frequency and nature of rupture-and-repair cycles evolve as users develop more sophisticated mental models of the AI, and how does this affect the emergence and stability of the Quasi-Creature perception?*
2. **Experimental Manipulation of the Uncanny Valley of Agency**: A series of controlled experiments could be designed to explicitly manipulate an AI's perceived agency (e.g., by varying the ratio of coherent to incoherent responses) to causally measure the effects on user frustration, trust, and cognitive load. The research question would be: *Can we identify a specific threshold of cognitive inconsistency that reliably plunges users into the uncanny valley of agency, and what design interventions can help mitigate this effect?*

3. **Investigating the Failed ToM Attempt**: Further research could use methods like think-aloud protocols or stimulated recall interviews to more deeply investigate the cognitive strategies users employ when their ToM for an AI fails. The research question would be: *What specific cognitive repair strategies (e.g., simplification, re-framing, abandonment) do users deploy when faced with an AI whose behavior violates their developing Theory of Mind?*
4. **Design Interventions for Graceful Failure**: Future work should explore design interventions aimed specifically at managing the Quasi-Creature perception and mitigating the associated frustration. This could involve designing systems that are more transparent about their limitations, that can signal when they have lost context, or that can engage in a rudimentary form of reason-giving for their outputs. This research trajectory could draw upon frameworks like Mutual Theory of Mind (MTOM), exploring how an AI that can build a simple model of its user's mental state might prevent or more gracefully recover from the interactional ruptures that trigger the frustrating Quasi-Creature perception (Wang et al., 2021).

Ultimately, learning to coexist with Quasi-Creatures requires us to move beyond the dream of creating a reflection of ourselves and to face the challenge of collaborating with an intelligence that is, and will likely remain, truly other. The relationship between humans and AI is co-evolutionary; a research agenda that embraces the complexity of human perception will be crucial for navigating this future responsibly and effectively.

communication between man and machine. *Communications of the ACM, 9*(1), 36–45.

Weizenbaum, J. (1976). *Computer Power and Human Reason: From Judgment to Calculation*. W.H. Freeman.

Wiener, N. (1948). *Cybernetics: Or control and communication in the animal and the machine*. MIT Press.

Winograd, T., & Flores, F. (1986). *Understanding computers and cognition: A new foundation for design*. Ablex Publishing.

Wobbrock, J. O., & Kientz, J. A. (2016). Research contributions in human-computer interaction. *ACM Interactions, 23*(3), 52–59.

Xiao, W., & He, C. (2020). Study on Anthropomorphism in Human-Computer Interaction Design. In S. Long & B. S. Dhillon (Eds.), *Man-Machine-Environment System Engineering. MMESE 2020. Lecture Notes in Electrical Engineering, vol 645*. Springer, Singapore. https://doi.org/10.1007/978-981-15-6978-4_72

Yampolskiy, R. V. (2024). *AI: Unexplainable, unpredictable, uncontrollable*. Chapman and Hall/CRC.

Yin, R. K. (2018). *Case study research and applications: Design and methods* (6th ed.). Sage publications.

Zambrano, R., Feger, S., Buschek, D., & Reiterer, H. (2024). Reproducibility Challenges with Large Language Models in HCI Research. *arXiv preprint arXiv:2404.15782*.

Zhang, A., & Varshney, L. R. (2025). Conceptualizing agency: A framework for human-AI interaction. In *CEUR Workshop Proceedings* (Vol. 3957, pp. 279–289). CEUR-WS.

Zuboff, S. (2019). *The age of surveillance capitalism: The fight for a human future at the new frontier of power*. PublicAffairs.